\documentclass[11pt]{article}
\pdfoutput=1
\usepackage{fullpage}
\usepackage{cite}
\usepackage{graphicx}
\usepackage{amsmath}
\usepackage{amssymb}
\usepackage{subcaption}
\usepackage{tcolorbox}
\usepackage{multirow}
\usepackage{microtype}
\usepackage{url}
\title{}
\date{}

\def\beq{\begin{equation}}
\def\eeq{\end{equation}}
\allowdisplaybreaks
\begin{document}
\bibliographystyle{utphys}
\newcommand{\msbar}{\ensuremath{\overline{\text{MS}}}}
\newcommand{\DIS}{\ensuremath{\text{DIS}}}
\newcommand{\abar}{\ensuremath{\bar{\alpha}_S}}
\newcommand{\bb}{\ensuremath{\bar{\beta}_0}}
\newcommand{\rc}{\ensuremath{r_{\text{cut}}}}
\newcommand{\Nd}{\ensuremath{N_{\text{d.o.f.}}}}

\titlepage
\begin{flushright}
QMUL-PH-21-14\\
\end{flushright}

\vspace*{0.5cm}

\begin{center}
{\bf \Large ``What happened, and who cared?''\\ 
Evidencing research impact retrospectively.}

\vspace*{1cm} \textsc{Chris
  D. White\footnote{christopher.white@qmul.ac.uk}, Anthony
  Phillips\footnote{a.e.phillips@qmul.ac.uk} and Beltran
  Sajonia-Coburgo-Gotha\footnote{b.sajonia-coburgogotha@se17.qmul.ac.uk}
} \\

\vspace*{0.5cm} Centre for Research in String Theory, School of
Physics and Astronomy, \\
Queen Mary University of London, 327 Mile End
Road, London E1 4NS, UK\\

\end{center}

\vspace*{0.5cm}

\begin{abstract}
Higher Education Institutions in the UK and elsewhere are under
increasing pressure to measure the impact of their research, which can
include how the research has increased scientific engagement amongst
the general public. For various reasons, the need for evidence can
arise months, or even years, after a particular research discovery has
been made. Furthermore, the right kind of evidence is needed to
indicate genuine behavioural change amongst a given target audience,
which can be difficult to obtain after time has passed. In this
article, we present a number of strategies for retrospective
evidencing of research engagement, and illustrate their use on
example discoveries from up to five years ago.
\end{abstract}

\vspace*{0.5cm}


\vspace*{0.5cm}

\begin{tcolorbox}
{\bf Key Messages}: 
\begin{itemize}
\item It is possible to evidence increased public engagement with
  scientific research months, or even years, after a discovery.
\item Systematic online data mining methods can be successfully
  employed, using freely available computational tools.
\item Different types of online data can be used for mutual
  corroboration in telling a convincing story e.g. social media data,
  comments from news articles or videos, and Wikipedia views.
\end{itemize}
\end{tcolorbox}

\section{Introduction}
\label{sec:intro}

Scientific research has long been recognised as a potentially valuable
contribution of the higher education sector. Recent years, however,
have seen increasing scrutiny of how to quantify this value. After
all, the financial resources available for science are necessarily
finite. Evaluating the relative benefits of different research
programmes can help in the allocation of these limited
funds. Furthermore, assessing the wider implications of research is a
way of systematising a moral obligation that many scientists feel:
that they should justify to funders -- including the taxpayer -- why
their research is important. 

The wider value of scientific research is usually referred to as {\it
  (research) impact}, and different definitions are adopted in
different academic contexts. Typically, however, one identifies one or
more {\it target audiences} outside academia, and then defines impact
as a {\it change in behaviour} of the audience(s). This is the
approach adopted by the UK Research Excellence Framework
(REF)~\cite{REF}, and an advantage of this definition is its
flexibility. For example, the impact may be commercial or
policy-based, where appropriate target audiences may include companies
or governmental organisations respectively. In this article, however,
we will be concerned with cultural or societal impact, and the target
audience we have in mind will be the general public. We may then talk
about the {\it reach} of any impact (how many people the research
influenced), and its {\it significance} (how deeply they were
affected). Whilst the former of these can be relatively easy to
demonstrate in large numbers (e.g. through audience figures for a
media outlet), the latter is notoriously difficult. 

For certain types of event (e.g., those falling under a traditional
public engagement banner, such as outreach talks or museum
exhibitions), one would usually include the evidencing of impact at
the project design stage. Examples include the use of before/after
questionnaires, or other similar means of gauging opinion, to
explicitly demonstrate that impact has occured. Indeed, this is
usually required by funding bodies. For instance, the UK Research
Councils have developed criteria for designing and evaluating public
engagement activies both individually (see, for instance, the STFC
Public Engagement Evaluation Framework~\cite{STFC_Public_engagement})
and through overarching bodies such as the National Co-ordinating
Centre for Public Engagement.  However, whilst the significance of
impact in such cases can often be well demonstrated, the reach tends
to be low.

An opposite case is that of individual research discoveries that are
shared widely by the media. Here the reach, as measured by viewing
figures or other metrics, can be enormous (the case studies presented
herein involve millions of people). However, this does not by itself
demonstrate any impact, as the latter requires an explicit
demonstration of behavioural change, which can be difficult to
achieve~\cite{Grant:2018:2399-8121:122}. There may be ways around this
-- such as looking at impact on science journalists themselves rather
than the public they broadcast to~\cite{Williams:2019:2399-8121:218}
-- but it remains desirable to consider the impact of research on
society as a whole. Compounding this problem is the fact that evidence
for such impact may need to be collected months, or even years, after
a particular discovery has been made, reasons for which include the
following:
\begin{itemize}
\item Reaction to a given research event may not be immediate, as it
  may take some time for its importance to be realised.
\item It may not be realised {\it a priori} that a given event will
  receive widespread attention, and thus mechanisms for recording its
  impact may not be put in place.
\item A given institution may lack resources or frameworks for
  systematically evidencing research impact as and when it occurs.
\item Changing external factors or assessment criteria for research
  institutions may create a need for evidence-gathering that was
  absent at the time of the discovery.
\end{itemize}
The question of how to retrospectively evidence research impact is
highly topical, given the growing impact agenda in the UK and
elsewhere.

A promising new avenue towards demonstrating research impact among a
broad public comes from the fact that we are living in a golden age of
data science. The ubiquity of computer resources and the internet make
potential sources of evidence both more varied and voluminous than
ever before. This already suggests that a number of different online
datasets might be useful for demonstrating impact, relating e.g. to
social media or news articles. Whilst some tools exist for analysing
such data, however, they are not necessarily geared towards evidencing
research impact, as ref.~\cite{Grant:2018:2399-8121:122} has recently
highlighted. The main reason for this is simply that specialist
academic knowledge of the research in question can be useful in
informing the design of suitable data-mining tools.

The aim of this paper is to demonstrate a number of ways that online
data mining can be used to evidence the impact of research
discoveries, where the latter may be a few years old. Our target
audience will be members of the general public, and the types of
behavioural change that we will seek to evidence include: (i)
widespread discussion of a particular discovery or research event;
(ii) increased engagement with a broad research area (e.g. astronomy)
following a specific event; (iii) increased understanding of research
topics or details. Importantly, we will try to tie these changes to
specific research papers or findings, which can be a requirement of
some impact assessment exercises~\footnote{The need to tie impact to
  specific research papers was a formal requirement in REF2014, but
  has been relaxed slightly (i.e. to allow reference to a body of work
  of associated individuals) in REF2021~\cite{REF}.}. 

The structure of our paper is as follows. In section~\ref{sec:online},
we outline the various online data sources that we will consider, and
describe the computational methods we have used to accrue sufficiently
large datasets for analysis. In section~\ref{sec:results}, we will
analyse the data we obtain in specific examples of fundamental physics
discoveries in the past few years, and present in each case examples
of behavioural change. We discuss our results and conclude in
section~\ref{sec:conclusion}.

\section{A survey of online data sources and ways to interact with them}
\label{sec:online}

The internet provides a large number of avenues for the general public
to engage in scientific research. Conversely, this allows academics
(or other research stakeholders) multiple entry points for evidencing
the societal impact of their results.

While each individual site offers unique ways to interact with its
material, a generic set of responses has become widely available,
which we summarise in Table~\ref{tab:engagement}. These responses
require different amounts of time (e.g., ``liking'' a post is faster
than writing a paragraph in response to it) and may be more or less
public (e.g., Twitter ``likes'' are publicly available, but
``retweets'' are explicitly drawn to the attention of all
followers). This might provide a helpful scale of engagement: responses
requiring greater time or social commitment (i.e., towards the bottom
right corner of Table~\ref{tab:engagement}) represent greater engagement.

In this section, we concentrate on a number of specific sources of
online data, and briefly describe how they can be used to construct
narratives of behavioural change.

\begin{table}
  \centering
  \begin{tabular}{llp{6cm}p{6cm}}
    &&\multicolumn{2}{c}{Social commitment} \\
    &&\textit{Low} & \textit{High} \\
    \hline
    \multirow{2}{*}[-3.5em]{\rotatebox{90}{Time commitment}} & \textit{Low} & \textbf{One-click reaction}
                                                            \par ``Liking'' material. Reactions may be displayed as a total number and/or used to sort comments. \par \textit{Examples:} Twitter ``likes'', Reddit ``up/downvotes'', news site reactions & \textbf{One-click recommendation} \par Sharing results with friends on a social network, without further comment. \par \textit{Examples:} Twitter ``retweets'', news site ``share'' buttons \\
    \cline{2-4}
    &\textit{High} & \textbf{Learning more} \par Following up references from media, either through explicit hyperlinks or by searching for keywords. \par \textit{Examples:} Wikipedia page visitor history & \textbf{Sharing knowledge} \par Initiating or contributing to ongoing discussion. \par \textit{Examples:} Twitter ``quote retweets'' or new tweets, Reddit replies or new posts, independent blog articles\\
    \hline
  \end{tabular}
  \caption{Schema of different generic ways to engage with online material on news sites or social media. We argue that, in general, greater time or social commitment both indicate greater engagement.}
  \label{tab:engagement}
\end{table}

\subsection{News articles}
\label{sec:news}

Whilst television remains the most popular source of news, its use is
falling. On the other hand, news consumption via the internet is
rising, with 66\% of all UK citizens aged 16+ relying on it in
2019. This outstrips similar figures for radio (43\%) and newspapers
(38\%). Furthermore, internet use dominates over television for two
(overlapping) demographics, namely young adults (aged 16-24), and
certain ethnic minorities~\cite{Ofcom}. From an impact point of view,
internet news has a distinct advantage over traditional TV or print
sources: many news websites allow users to post comments on any given
story. These comments may be supplemented by additional information,
such as a unique name or identifier for the author of each comment,
their geographical location, and a date or time stamp for the comment
itself. Some news websites allow users to rate comments using likes
and / or dislikes. Three of the most popular news websites in the UK
are the Mail Online (an offshoot of the Daily Mail newspaper), BBC
News and The Guardian (n.b. all of these outlets are free to consume,
with no registration required). Each of these three websites allow
user comments, although not necessarily on every story. Readership
figures and additional comment functionalities are summarised in
table~\ref{tab:news}.

Popular news stories can generate many (hundreds of) thousands of
comments, straddling a few years in some cases. It is clearly
inefficient to collect comments from large numbers of articles by
hand. However, the procedure can be automated using well-established
computational techniques. More specifically, we have written computer
codes in the \texttt{Python} language, that in turn rely on the
publicly available packages \texttt{Selenium} (for automated web
browsing) and \texttt{Beautiful Soup} (for parsing of website source code
in HTML). These can be used to efficiently strip comments from an
arbitrary news article associated with a given outlet, although custom
codes are required for each news source. For the present study, we
have manually searched for news articles relating to specific research
results, although this could also be automated in principle.

\begin{table}
\begin{center}
\begin{tabular}{c|c|c|c|c|c}
News outlet & Readership  & Date
\& Time & Location & Response\\ 
\hline \\
BBC & 35M & \checkmark &
$\times$ & $+/-$\\ 
Mail Online & 34M & \checkmark~$^*$ & \checkmark 
&$+/-$\\ 
The Guardian & 24M & \checkmark & $\times$ & $+$
\end{tabular}
\caption{Reach and comment functionality for three main UK news
  websites. Here the readership figures correspond to the total unique
  visitors / viewers in November 2019, and are taken from
  \texttt{comscore.com}. Also \checkmark~$^*$ denotes the fact that
  the Mail Online timestamp information becomes incomplete for older
  comments. Furthermore, $+$ and $-$ indicate whether positive or
  negative responses to comments are possible respectively.}
\label{tab:news}
\end{center}
\end{table}

There is clearly a large amount of information contained in news
comments. For example, numbers of comments on any given article, as
well as location information, can be used to estimate the
(inter)-national reach of a given discovery. Numbers of positive and
negative responses can evidence which particular aspects of a news
articles people are most strongly engaging with, and it may be
possible to relate such aspects to specific research papers. Of
course, a wealth of qualitative information is contained in the texts
of the comments themselves. One may look e.g. for prevalence of
keywords in order to ascertain if given topics are being talked about,
or evidence of {\it conversation} or {\it debate} involving two or
more commenters.

\subsection{Social Media}
\label{sec:socialmedia}

\begin{figure}
\begin{center}
\scalebox{0.6}{\includegraphics{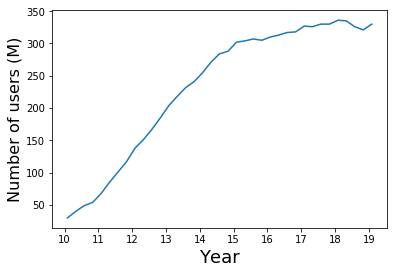}}
\caption{The number of active Twitter users as a function of
  time.}
\label{fig:twitter_users}
\end{center}
\end{figure}

Social media platforms such as Twitter and Facebook allow users to
share scientific information (including news articles), and to comment
on its significance or implications. Indeed, roughly half of UK adults
now use social media for news~\cite{Ofcom}. Here, we will focus on the
Twitter platform, whose number of active users has risen remarkably in
the past decade, remaining relatively stable from 2015 onwards (see
figure~\ref{fig:twitter_users}).
The Twitter website and app allows users to post short (140 character)
``tweets'' for consumption by their ``followers'', which may or may
not include additional media such as images or weblinks. Each tweet
carries the name of the tweeter (and their unique Twitter handle), as
well as date information. Users may reply to the tweets of others, and
also ``retweet'' them, so that they are shared for their own followers
to see. Finally, users can ``like'' the tweets of others, such that
any given tweet can be characterised by its number of replies,
retweets and (positive) reactions. 

Tweets may be mined retrospectively in a similar fashion to the news
comments of the previous section. Twitter itself provides dedicated
packages for Python, which can be used to systematically mine tweet
information in real time, or access past tweets. In particular, one
may search past tweets for certain phrases or keywords, including
``hash-tags'' (Twitter's own special keywords, which are always
preceded by the \# symbol). Examples of the use of Twitter data are
similar to that mentioned for news comments above. However, social
media has the advantage that the numbers of people engaging with a
given tweet or topic can be much larger: we will see examples
involving hundreds of thousands of people in what follows.

\subsection{YouTube}
\label{sec:youtube}

YouTube is a global video sharing platform, with over 1.9 billion
logged in users per month. Anyone can register to upload videos on the
site, and the ubiquity and on-demand nature of its content (70\% of
which is viewed on mobile phones) means that it is replacing
traditional TV-based science content for many people. Examples of
research-based videos considered here include those created by public
organisations (e.g. NASA), journalists, and dedicated YouTube science
channels. On each YouTube video page, one may easily see the number of
views a video has had, and a number of positive and negative reactions
(analogous to the likes and dislikes in news articles). Viewers may
comment on videos, and the sheer number of viewers involved (sometimes
several millions) means that a large amount of comment data can be
mined, using similar methods to those described above. Each comment
has a number of ``votes'', which plays a similar role to the number of
positive reactions for news comments. One may also obtain date
information for each comment, although this becomes less precise for
older comments.

\subsection{Wikipedia}
\label{sec:wiki}

Wikipedia is a free online encyclopedia, that anyone can edit. It has
rapidly beome a goto online resource, replacing traditional print
encyclopedias, particularly given its dynamic nature that can respond
to events and discoveries in real time. It attracts hundreds of
millions of monthly visitors, and contains roughly 50M articles, in
302 languages. Web pages exist for specific topics, and are also
linked together so that one can easily view pages which are closely
related to a given article. Data on Wikipedia usage, including page
views in different languages and / or as a function of time, can be
easily exported from the publicly available tool \texttt{PageViews
  Analysis}. Examples of how to use this data include evidencing the
reach of public impact associated with a particular discovery, which
is particularly straightforward if a dedicated Wikipedia article
exists. Another very useful idea is to look at correlations between
different articles: one may try to argue that people who engage with a
particular discovery or research result are in turn more likely to
read more widely about the underlying subject and context of the
research. This narrative becomes particularly powerful if combined
with data from one of the alternative sources outlined in the previous
sections.

\subsection{Reddit}
\label{sec:reddit}

Reddit is a news aggregation and discussion website, that proudly
calls itself ``the front page of the internet''. Users can post texts,
web links (e.g. to news articles) or images, which are grouped into
subject-specific forums or {\it subreddits}. Each post can be voted up
or down by website members, so that individual posts may move to the
top of a given subreddit, or even appear on the main website
itself. Reddit is the 18$^{\rm th}$ most visited website in the world,
with 42-49.3\% of users originating from the US, and 7.9-8.2\% from
the UK. There are a number of insights that can be obtained from
analysing Reddit data. Firstly, there are quantitative metrics such as
the number of posts relating to a given research discovery as a
function of time, and the number of upvotes received. Such information
can be used alongside similar metrics from analysing Wikipedia and
Twitter data, in order to build a consistent narrative of public
engagement. One may also look for use of specific images relating to a
given research project, and measure the proportion of posts which
feature them. Secondly, there are valuable qualitative details that
can be gathered from Reddit data. Each subreddit corresponds to a
distinct community of people, united by a special interest. The
different subreddits that discuss a particular research discovery then
provide a cross-section of which societal groups have been
enthused. Not only can this be used to classify the impact of a
particular discovery, but it can also inform the design of future
impact strategies. 

In summary, we have described a number of different avenues for
collecting online data related to public engagement with scientific
research. Each of these in isolation tells an incomplete story, and
must be used with care. In particular, one has little or no knowledge
of the demographic associated with each online tool. However, by
combining information from different sources, one can start to build
up a more complete picture of how a given set of research results has
entered the public conciousness, and generated discussion and / or
debate. Let us now begin to do this, by focusing on specific examples.

\section{Examples of specific research projects}
\label{sec:results}

\subsection{The discovery of Proxima b}
\label{sec:proximab}

A significant focus of astronomical research in recent years has been
the search for {\it exoplanets}, namely planets which orbit stars
other than our own Sun. As well as the intrinsic interest in
cataloguing the presence of celestial bodies outside our solar system,
the study of exoplanets may yet reveal the existence of
extra-terrestrial life. Discovery of the latter would lead to a
profound reevaluation of our place in the universe, and have
far-reaching implications for the philosophy and religious faiths of
humans around the globe. In 2016, the {\it Pale Red Dot} project
discovered an exoplanet orbiting our nearest star, Proxima Centauri,
which itself forms part of the Alpha Centauri star
system~\cite{2016Natur.536..437A}. This remains the nearest exoplanet
to Earth ever discovered. Furthermore, ongoing studies of its
environment suggested that it may harbour conditions for
extraterrestrial life. These facts led to considerable media reporting
and attention worldwide, including print, broadcast and online
media. As discussed above, however, this does not by itself constitute
impact of the discovery, given that merely encountering a news article
does not necessarily amount to behavioural change of the observer. In
order to measure the latter, we have used the computational methods
discussed in section~\ref{sec:online} to construct a database of over
57.5k tweets relating to ``Proxima Centauri''. Our use of this
particular search term is motivated by its being the parent star about
which the newly discovered Proxima b orbits. Discovery of the star
predates the discovery, such that we would expect increased discussion
relating to Proxima Centauri to correlate with key dates in the
discovery timeline of Proxima b. We have also accrued 4910 news
comments (from the Daily Mail, Guardian and BBC websites); 1583 Reddit
posts relating to Proxima b from 484 distinct subreddits; and
Wikipedia information using PageViews Analysis.  This allows us to
provide evidence for an increased public engagement in astronomy
topics relating to exoplanets, Proxima Centauri and the likelihood of
extraterrestrial life.

\begin{figure}
\begin{center}
\begin{subfigure}{0.4\textwidth}
\centering
\includegraphics[width=\textwidth]{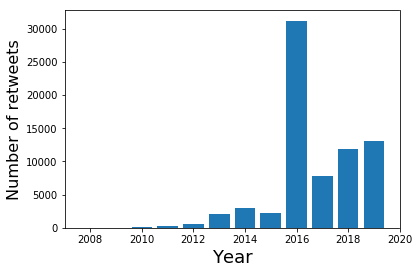}
\caption{}
\end{subfigure}
\begin{subfigure}{0.4\textwidth}
\centering
\includegraphics[width=\textwidth]{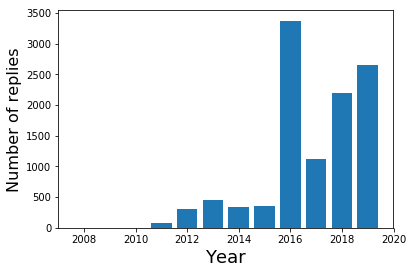}
\caption{}
\end{subfigure}
\end{center}
\caption{(a) The number of retweets per year relating to the search
  term ``Proxima Centauri''; (ii) the number of Twitter replies.}
\label{fig:proxima_twitter}
\end{figure}

In figure~\ref{fig:proxima_twitter}(a), we show the number of retweets
which contain the term ``Proxima Centauri''. We choose retweets rather
than tweets, as these indicate a more active engagement of twitter
users, who have thus explicitly chosen to rebroadcast a particular
piece of information. Likewise, in
figure~\ref{fig:proxima_twitter}(b), we show the number of replies to
tweets featuring the same search term. Both plots start with a gentle
rise in discussion, which comparison with
figure~\ref{fig:twitter_users} reveals is not due to to increased
public engagement with astronomy, but rather the increase in the
number of global Twitter users. Nevertheless, the most dramatic
feature in both figures~\ref{fig:proxima_twitter}(a) and (b) is a
highly pronounced spike in activity coinciding with the discovery year
(2016) of Proxima B. The number of global retweets for the selected
search term was over 23 times higher than the four-year average
pre-discovery, and the annual average post-discovery is almost 8 times
the average pre-discovery. This is far in excess of the rise
in the number of Twitter users in this period. Furthermore, the trend
in engagement post-discovery is upward, indicative of a sustained
impact, as Proxima b enters wider popular culture, with increased
media references.

\begin{figure}
\begin{center}
\scalebox{0.6}{\includegraphics{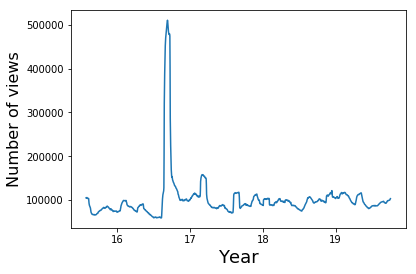}}
\caption{Daily viewing figures for Wikipedia articles on {\it Proxima
    Centauri} in 73 different languages, where the curve has been
  smoothed with a 30-day rolling average.}
\label{fig:proxima_wikiviews}
\end{center}
\end{figure}

A similar story is told by the Wikipedia data:
figure~\ref{fig:proxima_wikiviews} shows the number of views of the
Wikipedia pages for {\it Proxima Centauri} (in 73 different languages)
as a function of time. A significant spike is seen around discovery,
and a sustained increase in activity occurs afterwards. To quantify
this, we may avoid the spike by calculating the average number of
views after $1^{\rm st}$ October 2016, and before $1^{\rm st}$ August
2016. We then find an average increase in engagement of 30\%,
amounting to 20k extra views per month.
On Reddit, there were over 350 posts on the discovery date itself, and
there has been a sustained activity since, with around 4.7 posts per
week on average. 

\begin{figure}
\begin{center}
\begin{subfigure}{0.4\textwidth}
\centering
\includegraphics[width=\textwidth]{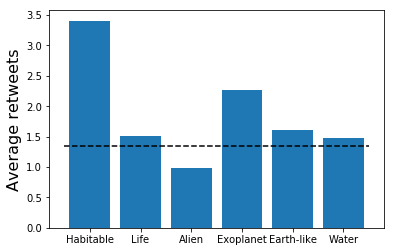}
\caption{}
\end{subfigure}
\begin{subfigure}{0.4\textwidth}
\centering
\includegraphics[width=\textwidth]{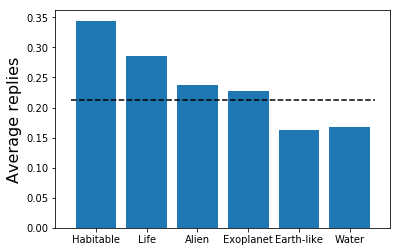}
\caption{}
\end{subfigure}
\end{center}
\caption{(a) The average number of retweets for tweets containing
  ``Proxima Centauri'', containing certain keywords; (b) Similar but
  for replies. In each case, the dashed line indicates the average
  activity for all ``Proxima Centauri'' tweets.}
\label{fig:proxima_keywords}
\end{figure}

\begin{figure}
\begin{center}
\scalebox{0.6}{\includegraphics{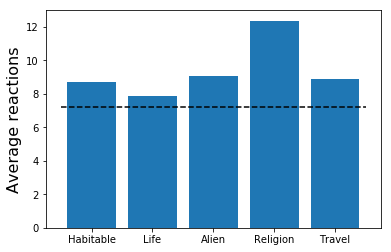}}
\caption{Average number of reactions for news comments containing
  certain keywords, where the dashed line denotes the average for all
  comments in the selected articles.}
\label{fig:proxima_news}
\end{center}
\end{figure}

The quantitative results above show increased engagement of the public
with astronomy related to the discovery of Proxima b. Further valuable
insights on the nature of the impact can be gained by examining
qualitative data. First, we may perform a keyword analysis on the
collected texts in our database of
tweets. Figure~\ref{fig:proxima_keywords}(a) shows the average number
of retweets and replies for tweets containing certain
keywords. Depending on which measure is used, above-average activity
is strongly associated with the concept of habitability. Our other
chosen keywords demonstrating above-average engagement are clearly
related to the potential existence of extra-terrestrial life, and we
found that the number of tweets containing the words ``habitable'',
``water'', ``alien'' or ``life'' strongly increased after the 2016
discovery, an increase that continued thereafter. Similar results
arise from analysing our collected news comments:
figure~\ref{fig:proxima_news} shows the average number of reactions
for news comments containing various keywords. As well as
corroborating the interest in alien life, above-average engagement is
also observed for comments discussing the implications of the
discovery for religion, as well as the possibility of interstellar
travel to the exoplanet (the latter was well-publicised after a press
conference for the {\it Breakthrough Foundation}, who initiated a
project to send small spacecraft to Proxima b).
This consistency of popular keywords across different media platforms
significantly supports the hypothesis that public interest in the
Proxima b discovery focuses on its implications for the existence of
life outside our solar system.

\begin{figure}
\begin{center}
\begin{subfigure}{0.4\textwidth}
\centering
\includegraphics[width=\textwidth]{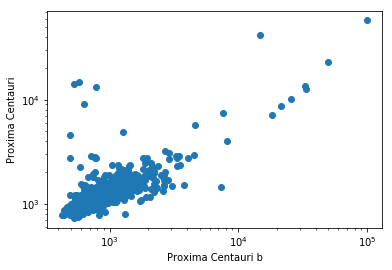}
\caption{}
\end{subfigure}
\begin{subfigure}{0.4\textwidth}
\centering
\includegraphics[width=\textwidth]{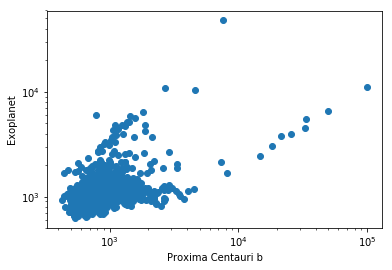}
\caption{}
\end{subfigure}
\end{center}
\caption{Viewing figures for Wikipedia articles, plotted against the
  viewing figures for the article {\it Proxima Centauri b}, taken on
  the same day.}
\label{fig:proxima_correlations}
\end{figure}

The Twitter data already provides evidence that more people are
finding out about Proxima Centauri as a result of the Proxima b
discovery. It is also possible to show this also using the Wikipedia
data. In particular, one may look at {\it correlations} between
viewing figures for different Wikipedia articles. {\it Proxima
  Centauri b} (to use the full name of the exoplanet) has had its own
Wikipedia article since $24^{\rm th}$ August 2016. One may then plot
the viewing figures for related articles on a particular day, vs. the
viewing figures for the Proxima b article~\footnote{Given that the
  Proxima b article appeared originally only in English, we have
  chosen to compare correlations with related articles in a single
  language.}. We show this in figure~\ref{fig:proxima_correlations}
for the articles on {\it Proxima Centauri} and {\it Exoplanets}. There
is a clear correlation in each case, suggesting that people who are
reading about Proxima b are in turn seeking out more information on
the host star, or on Exoplanet research in general. The positive
correlation can be quantified in each case by the well-known {\it
  Pearson correlation coefficient}, which is one for complete
(positive) correlation, and zero for no correlation. The Pearson
coefficients are 0.84 and 0.30 for the Proxima Centauri and Exoplanet
articles respectively, with negligible uncertainties~\footnote{One may
  worry that the statistical assumptions underlying the use of the
  Pearson coefficient - that of a linear relationship, and Gaussian
  uncertainties for the viewing figures - do not apply here. An
  alternative is to use the {\it Spearman coefficient}, which
  evaluates to 0.72 and 0.37 for the two cases.}. Correlation does not
imply causation in general, and there are known statistical problems
when comparing time series of data, as we are doing here. However, the
fact that the Wikipedia correlation agrees with similar conclusions
reached from the social media data, means that one can be confident
that a genuine correlation is being observed.

In addition to the keyword analysis above, one may examine which
groups of people were particularly influenced by the Proxima b
discovery. The Reddit communities with the highest proportion of
Proxima b posts include one focused on creative writing and another
dedicated to “exhibiting the awesome potential of humanity”,
indicating that Proxima b has uniquely captured public imagination
beyond typical astronomy-interested audiences.

\subsection{The Cassini-Huygens mission}
\label{sec:cassini}

As a second example of a research with a significant public engagement
component, we use the results of the {\it Cassini-Huygens probe}. This
was a NASA satellite mission, which reached the planet Saturn in
2004. It then spent 13 years orbiting the planet, performing detailed
observations of its moons and ring systems, including Enceladus and
Titan, which are thought to be potentially habitable by microbial
extraterrestrial life. In September 2017, the mission culminated in a
spectacular {\it Grand Finale} phase, in which the probe was destroyed
by intentionally crashing into Saturn, in order not to contaminate the
nearby moons. During its lifetime, the Cassini-Huygens mission
captured over 400k images of Saturn and its environs. These generated
new research results in planetary mechanics, but also gained
widespread public interest. One picture in particular - dubbed
``Cassini's most iconic image mosaic'' by NASA Project Scientist Linda
Spilker - featured a high-resolution shot of Saturn, with the Earth as
a tiny speck in the background. The striking poignancy of the image
was highlighted by news outlets around the world. Likewise, the Grand
Finale phase of the mission also received widespread media attention.

Similar to the exoplanet discovery discussed in the previous section,
the Cassini mission is able to potentially stimulate public interest
in astronomy. It may also have changed the status quo regarding how
society at large both visualises and understands Saturn. Proving this
retrospectively, however, poses different challenges to the previous
example. Despite the fact that iconic images have been produced -
whose (re-)use in online media can potentially be traced - there is
not necessarily a clear demarcation between ``before'' and ``after''
when it comes to discussing Saturn. In the exoplanet example, {\it
  Proxima b} did not exist before its discovery, and so any discussion
of its properties constitutes a shift in public understanding. This is
not the case with Saturn, such that different ways of thinking about
its impact are needed. Nevertheless, similar techniques may be used to
the previous section, in building a coherent narrative for public
behavioural change. To this end, we have amassed 223k tweets relating
to the search string ``Saturn Cassini'', from 2013 onwards; 6080
comments from news articles relating to Cassini on the Daily Mail,
Guardian and BBC websites; 12,300 comments from YouTube videos
relating to Cassini; over 125k Reddit posts relating to Saturn in over
12k distinct subreddits; and Wikipedia pageview data using PageView
Analysis.

\begin{figure}
\begin{center}
\begin{subfigure}{0.4\textwidth}
\centering
\includegraphics[width=\textwidth]{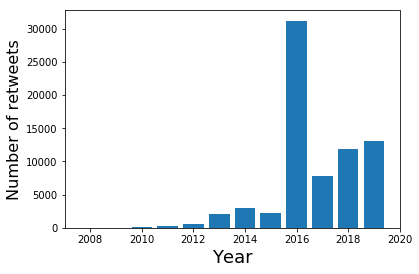}
\caption{}
\end{subfigure}
\begin{subfigure}{0.4\textwidth}
\centering
\includegraphics[width=\textwidth]{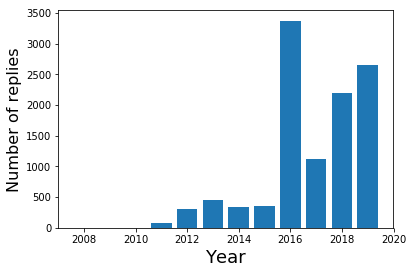}
\caption{}
\end{subfigure}
\end{center}
\caption{(a) The number of retweets per year relating to the search
  term ``Saturn Cassini''; (ii) the number of Twitter replies.}
\label{fig:cassini_twitter}
\end{figure}

\begin{figure}
\begin{center}
\scalebox{0.2}{\includegraphics{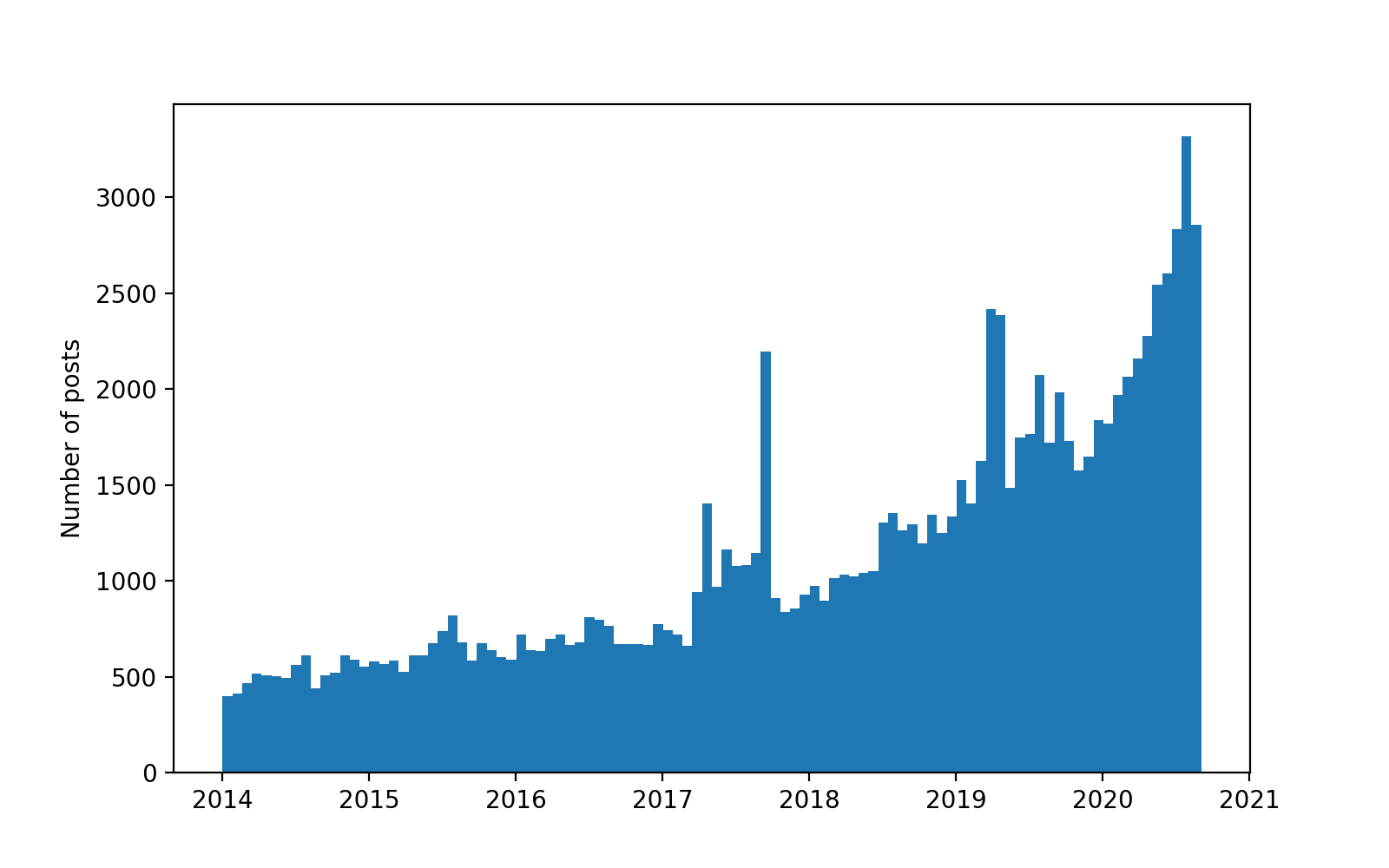}}
\caption{Rise in Reddit activity related to Saturn.}
\label{fig:saturn_reddit}
\end{center}
\end{figure}

\begin{figure}
\begin{center}
\scalebox{0.6}{\includegraphics{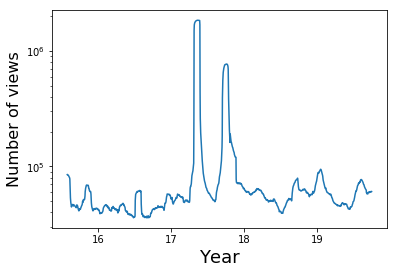}}
\caption{Daily viewing figures for Wikipedia articles on {\it
    Cassini-Huygens} in 72 different languages, where the curve has
  been smoothed with a 30-day rolling average.}
\label{fig:cassini_wikiviews}
\end{center}
\end{figure}

Quantitative evidence for a sustained increase in public engagement
with the mission is revealed in the Twitter data. In
figure~\ref{fig:cassini_twitter}, we show the number of retweets and
replies to tweets relating to ``Saturn Cassini'', which demonstrate a
clear peak in engagement throughout the Grand Finale year e.g. there
were over 287k retweets in 2017, 9.5 times higher than the mean in the
preceding three years. There is also a sustained increase thereafter,
showing widespread engagement with the mission's legacy. Our Reddit
data also shows a sustained increase in engagement with Saturn-related
content, with weekly posts rising from a baseline of around 200 to
around 800 after the peak in 2017 (figure~\ref{fig:saturn_reddit}),
with a notable rise thereafter.
Wikipedia data confirm this picture. We found, for example, that page
views for Wikipedia's articles on Saturn spiked by approximately a
factor of 9 during the Grand Finale orbit. In
figure~\ref{fig:cassini_wikiviews}, we show the viewing figures for
articles on the Cassini-Huygens mission (in 72 different
languages). The two large spikes correspond to the start and end of
the Grand Finale orbit, which both received major press attention.
One expects, in the case of Cassini, that it has set a new paradigm in
how we visualise Saturn, in thats images have become definitive, even
amongst the general public. This can be evidenced by looking at our
Reddit data, and in particular the number of image-related posts about
Saturn. In a sample of 5645 posts, we found that 18\% feature Cassini
images, showing that significant numbers of people are choosing Cassini
results to illustrate their discussions.

\begin{figure}
\begin{center}
\begin{subfigure}{0.4\textwidth}
\centering
\includegraphics[width=\textwidth]{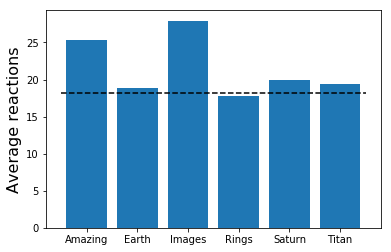}
\caption{}
\end{subfigure}
\begin{subfigure}{0.4\textwidth}
\centering
\includegraphics[width=\textwidth]{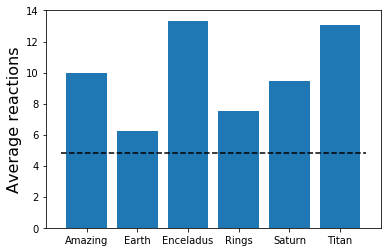}
\caption{}
\end{subfigure}
\end{center}
\caption{(a) The average number of reactions for comments relating to
  the Cassini mission, from (a) news articles; (b) YouTube
  videos. Also shown is the average number of reactions across all
  comments.}
\label{fig:cassini_keywords_news}
\end{figure}

As for the exoplanet discovery discussed above, one may perform a
keyword analysis to see which aspects of Cassini science are the
drivers of public engagement. Interest in the moons (e.g. Titan) is
evident, as are new results on the ring system. The keyword ``Earth''
gets the most attention, which is tied to the iconic {\it Day the
  Earth Smiled} image mentioned above. Interestingly, engagement as
measured by {\it replies} to tweets (perhaps a more active measure of
engagement) is significantly higher than average, as compared to
retweets. This story is replicated by reaction data taken from news
article comments, and from YouTube videos
(figure~\ref{fig:cassini_keywords_news}). In the latter case, interest
can also be seen in Enceladus (another of Saturn's moons). Comments
which explicitly talk about the wonder conveyed by Cassini images (as
measured by the word ``Amazing'') also lead to increased engagement.
\begin{figure}
\begin{center}
\begin{subfigure}{0.4\textwidth}
\centering
\includegraphics[width=\textwidth]{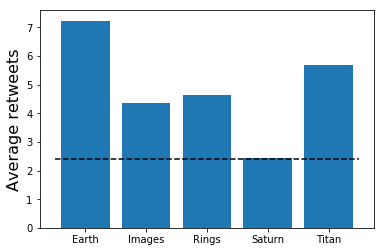}
\caption{}
\end{subfigure}
\begin{subfigure}{0.4\textwidth}
\centering
\includegraphics[width=\textwidth]{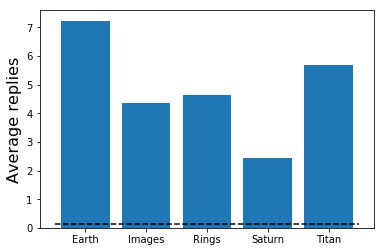}
\caption{}
\end{subfigure}
\end{center}
\caption{(a) The average number of retweets for tweets containing
  ``Saturn Cassini'', containing certain keywords; (b) Similar but
  for replies. In each case, the dashed line indicates the average
  activity for all ``Saturn Cassini'' tweets. }
\label{fig:cassini_keywords}
\end{figure}

\begin{figure}
\begin{center}
\begin{subfigure}{0.4\textwidth}
\centering
\includegraphics[width=\textwidth]{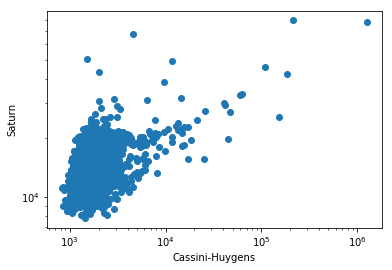}
\caption{}
\end{subfigure}
\begin{subfigure}{0.4\textwidth}
\centering
\includegraphics[width=\textwidth]{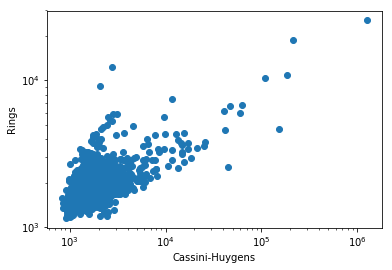}
\caption{}
\end{subfigure}
\end{center}
\caption{Viewing figures for Wikipedia articles, plotted against the
  viewing figures for the article {\it Cassini-Huygens}, taken on
  the same day.}
\label{fig:cassini_correlations}
\end{figure}

Evidence for the public educating themselves can again be gleaned by
looking at correlations of Wikipedia page views. In
figure~\ref{fig:cassini_correlations}, we show the page views for
related articles to the Cassini-Huygens article (across all
languages), compared with the latter article itself. There is a clear
correlation with people reading about Saturn in general, or the ring
system: the respective Pearson correlation coefficients are 0.43 and
0.70, with negligible uncertainties~\footnote{The respective Spearman
  correlation coefficients are 0.35 and 0.39, again with negligible
  uncertainties.}. 
As for our previous research example, the fact that the keyword
analysis tells a similar story of engagement means that we can be
confident that a genuine correlation is being observed. 

Further qualitative insights can be gained by looking at which Reddit
subforums users are engaging with. We found that the top 20 subreddits
for Cassini-related discussions include nine focused on general
photography and general interest posts. In 2019-2020 there was an
average of five posts a week, even on advanced topics such as the moon
Enceladus. Thus, the image-related aspect of the Cassini mission has
indeed informed discussions in communities outside of typical science
channels. Unsurprisingly, Reddit posts with the most upvotes convey
inspiration and wonder; we found multiple posts with more than 10,000
upvotes and / or more than 1,000 comments. Qualitative information
from comments can be highly useful. In our database of combined news
and YouTube comments, for example, we found evidence of the following:
\begin{itemize}
\item {\it Increased understanding} of space exploration e.g. ``“\$4bn
  well spent - the knowledge Cassini has provided has broadened our
  understanding of our Solar System and hopefully paved the way for
  future manned exploration.'' (a news comment with 342 reactions).
\item {\it Specific references to scientists} from particular
  institutions (n.b. this can be useful for assessment exercises such
  as the REF).
\item {\it Debate and discussion} about complex scientific topics. In
  the Cassini case, the YouTube comments show people discussing the
  nature and structure of Saturn's rings (e.g. ``Saturn has moons
  between its rings that produce ripples in the rings through
  gravity...interesting!''); why the Cassini mission flew into Saturn
  and the proof that it did so; and the nature of the instrumentation
  on board, and how this influenced the reported images
  (e.g. ``Cassini's speed is the reason these particular photos are
  only in greyscale and low resolution on this pass'').
\end{itemize}

\section{Conclusion}
\label{sec:conclusion}

In this paper, we have examined the problem of how to evidence the
impact of well-publicised research discoveries on the general public,
years after a given discovery took place. By exploiting the wealth of
data available online, we have constructed narratives of increased
public engagement for two example high-profile research projects: (i)
the discovery of the exoplanet Proxima b in 2016; (ii) the
Cassini-Huygens mission, which culminated in a spectacular Grand
Finale in 2017. We have constructed our own computational tools for
extracting and analysing Twitter data, Reddit posts, Wikipedia views,
and news / YouTube comments. This is greatly advantageous to using
pre-existing online metrics or tools, which would need to be tailored
to be able to incorporate specialist research knowledge related to the
projects at hand.

Whilst each data type by itself may provide relatively weak evidence
for behavioural change, the combination of several mutually
corroborative elements leads to much stronger conclusions. We found
convincing evidence for increased public awareness of astronomy
topics; discussion and debate of advanced scientific concepts; and a
sustained legacy of engagement with our chosen research results. This
suggests that similar methods may prove useful in future, particularly
for the kind of fundamental physics discoveries that - whilst highly
publicised - can be tricky to describe in terms of the impact
narratives required by assessment exercises such as the REF.

Our paper is very much a proof-of-concept study, and there are clearly
many avenues for future research. The investigation of research
projects which are not quite as high-profile as the ones presented
here would be interesting, as would the extension of our datasets to
include more types of online data. It would also be possible to do a
much more sophisticated analysis of our comment databases. One might
think, for example, of applying sentiment analysis, natural language
processing (or other machine learning) techniques in order to quantify
how the public's response to scientific content changes over time, or
to quantify evidence of rigorous debate. We leave the investigation of
these interesting questions to future work, noting that the increasing
impact agenda, together with the rapid development of data science,
provide ample scope for interdisciplinary collaboration in this
area.

\section*{Acknowledgments}
We thank Guillem Anglada-Escud\'{e}, Clark Baker, Gavin Coleman,
Christophe Eizaguirre, Tom Horner, Carl Murray, Richard Nelson, Kati
Schwab, John Strachan and Natalie Wall for useful discussions.

\bibliography{refs.bib}
\end{document}